\documentstyle[12pt]{article}
\begin{document}
\large
\begin{center}
{\bf Impossible Realization of
Neutrino Resonance Oscillations Enhancement
in Matter}\\
\vspace{1cm}
Khamidbi M. Beshtoev \\
\vspace{1cm}
Joint Institute for Nuclear Research, Joliot Curie 6\\
141980 Dubna, Moscow region, Russia\\
\vspace{1cm}
\end{center}
\par
{\bf Abstract}
\par
In this paper, we study a photon, a massive charged particle, and a
massive neutrino passing through matter.
The hypothetical left-right symmetric weak interaction, which is used
in Wolfenstein's
equation can generate the resonance enhancement of
neutrino oscillations in matter, which disappears when
neutrinos go out into vacuum from matter (the Sun).
It is shown that
since standard weak interactions cannot generate masses, the laws of
conservation of the energy and the momentum of neutrino in matter will be
fulfilled only if the energy $W$ of polarization of matter by
the neutrino or the
corresponding term in Wolfenstein's equation, is zero.
This result implies that
neutrinos cannot generate permanent polarization of matter.  This
leads to the
conclusion:  resonance enhancement of neutrino oscillations in matter does
not exist.
It is also shown that in standard weak interactions the
Cherenkov radiation cannot exist.\\


\section{Introduction}
\par
In previous works [1], we studied the physical foundations
of Wolfenstein's equation [2], resulting in the
conclusion: in
Wolfenstein's equation, a hypothetical left-right symmetrical weak
interaction is used, but not the standard (left-side) weak interaction.
Therefore,
the resonance enhancement of neutrino oscillations in matter
obtained in Ref [2] have no connection with the weak interaction
 physics.
\par
This work continues the discussion on the problem of neutrinos passing
through matter.

\section{Wolfenstein's Equation for Neutrino in Matter
and the Mechanism of Resonance Enhancement of Neutrino Oscillations
in the Framework of a Hypothetical Weak Interaction}

\par
In the ultrarelativistic limit, the evolution equation for
the neutrino wave function $\nu_{\Phi} $  in
matter has the form [2]
\par
$$
i \frac{d\nu_{\Phi}}{dt} = ( p\hat I +
\frac{ {\hat M}^2}{2p} + \hat W ) \nu_\Phi ,
\eqno(1)
$$
\par
\noindent
where
$p, \hat M^{2}, \hat W $ are, respectively, the momentum,
the (nondiagonal) square mass matrix in
vacuum, and  the  matrix  taking  into account
neutrino interactions in matter,
$$
\nu_{\Phi} = \left (\begin{array}{c} \nu_{e}\\
\nu_{\mu} \end{array} \right) ,
\qquad
\hat I = \left( \begin{array}{cc} 1&0\\0&1 \end{array} \right) ,
$$
$$
\hat M^{2} = \left( \begin{array}{cc} m^{2}_{\nu_{e}\nu_{e}}&
m^{2}_{\nu_{e} \nu_{\mu}}\\ m^{2}_{\nu_{\mu}\nu_{e}}&
m^{2}_{\nu_{\mu} \nu_{\mu}} \end{array} \right).
$$
\par
As we can see from the form of Eq. (1), the above equation holds
the left-right symmetric neutrinos wave function
$\Psi(x) = \Psi_L(x) + \Psi_R(x)$.
This equation contains the term $W$, which arises from the weak
interaction (contribution of $W$ boson) and contains only a left-side
interaction of the neutrinos, and is substituted in the left-right
symmetric equation (1) without indication of its left-side origin.
Then we see that equation (1) is an equation that includes term $W$
which arises not from the weak interaction but from a hypothetical
left-right symmetric interaction (see  also works [1, 3]).
Therefore this equation is not one for neutrinos passing
through real matter. The problem of passing neutrinos through real matter
will be discussed in the next section.
\par
The matrix $\hat M^{2}$ is diagonalized by rotation through the
angle $\theta $ ($\theta$ is the angle of vacuum oscillation):
$$
\tan(2\theta) = \frac{2 m^{2}_{\nu_{e}\nu{\mu}}}{\mid m^{2}_{\nu_{\mu}
\nu_{\mu}} -
m^{2}_{\nu_{e}\nu_{e}} \mid},  \hspace{1cm}
\hat M^{2}_{diag} = \left( \begin{array}{cc}
m^{2}_{1}&0\\0&m^{2}_{2} \end{array} \right),
\eqno(2)
$$
$$
m_{1,2}^{2} = \frac{1}{2} [(m_{\nu_{e}\nu_{e}}^{2} +
m_{\nu_{\mu}\nu_{\mu}}^{2}) \pm \sqrt{(m_{\nu_{e}\nu_{e}}^{2} -
m_{\nu_{\mu}\nu_{\mu}}^{2})^{2} + 4 m_{\nu_{e}\nu_{\mu}}^{4}}] ,
$$
$$
\Delta m^{2} = \sqrt{(m_{\nu_{e}\nu_{e}}^{2} -
m_{\nu_{\mu}\nu_{\mu}}^{2})^{2} + 4 m_{\nu_{e}\nu_{\mu}}^{4}} ,
$$
and the length of vacuum oscillation    $L_{0}$
is
$$
L_{0} = \frac{4 \pi p}{\mid m^{2}_{1} - m^{2}_{2} \mid} ,
\hspace{1cm} E \cong pc.
\eqno(3)
$$
Since $\hat M^{2} $ is a nondiagonal matrix,
this vacuum oscillation of neutrinos will take place
at any energies with the length of oscillation $ L_{0} $.
\par
The solution of equation  (1), i.e. one for passing neutrinos through
a "matter" where they participate in a left-right symmetric hypothetical
weak interaction, is considered in detail in [4]
and here the main results will be shown in the reduced form.
\par
When neutrinos are passing through  "matter", their  influence (see equation
(1))
leads to changes of the  rotation angle $\theta $ for diagonalizing
the mass matrix $ \hat M^{2}$,
if diagonal  matrix $\hat W$,  responsible
for
the difference between the interactions of the neutrinos $(\nu _{e}, \nu
_{\mu })$,
is added to the mass term
$\hat M^{2}/2p$, and then $\theta$ becomes $\theta '
(\theta ' \neq  \theta )$.
\par
Thus, neutrino mixing in "matter" is determined by $\sin^{2}(2\theta')$:
$$
\sin^{2}(2\theta') = \sin^{2}(2\theta)/ [(\cos(2\theta) - \frac{L_{0}}
{L^{0}})^2 + \sin^{2}(2\theta)],
\eqno(4)
$$
$$
m_{1,2}^{'2} = \frac{1}{2} [(m_{\nu_{e}\nu_{e}}^{'2} +
m_{\nu_{\mu}\nu_{\mu}}^{'2}) \pm \sqrt{(m_{\nu_{e}\nu_{e}}^{'2} -
m_{\nu_{\mu}\nu_{\mu}}^{'2}) + 4 m_{\nu_{e}\nu_{\mu}}^{4}}] ,
$$
$$
\Delta m^{'2} = \sqrt{(m_{\nu_{e}\nu_{e}}^{'2} -
m_{\nu_{\mu}\nu_{\mu}}^{'2})^{2} + 4 m_{\nu_{e}\nu_{\mu}}^{4}} ,
$$
$$
L'_{0} = \frac{L_{0}}{\sin(2\theta)} ,
$$
where $m'_{\nu_{e}\nu_{e}}, m'_{\nu{\mu}\nu_{\mu}}$ are masses of
$\nu_{e}, \nu_{\mu}$ in "matter",
$L^{0}$ is a diffraction length (i.e., length of formation),
$$
L^{0} =2\pi m_{p}( 2^{0.5} G_{F} \rho Y_{e})^{-1} =
$$
$$
3 {10}^{7} (m) (\rho(g/cm^{3}) 2Y_{e})^{-1},
\eqno(5)
$$
\par
\noindent
$Y_{e}$ is the number of electrons per nucleon in "matter".
\par
In the common case $\theta '$  depends on the difference of masses
$m_{1}, m_{2}$, density $\rho$ of "matter" and the neutrino momentum.
\par
At $L'_{0} \cong L^{0}$ the resonant neutrino oscillations take place,
i.e.,  $\sin^{2}(2\theta')\cong 1$ or $\theta' \cong \frac{\pi}{4}$.

But, since $\Delta m^{'2} \cong m_{\nu_{e}\nu_{\mu}}^{2}$, the length of
oscillations in the "matter"  $L'_{0}$ is defined by
$m_{\nu_{e}\nu_{\mu}}^{2}$
$$
L'_{0} = \frac{4 \pi p}{\mid m_{\nu_{e}\nu_{\mu}}^{2} \mid}
$$
and increases relatively to the vacuum oscillations length $L_{0}$.
\par
We remind that at resonance (if $\nu_{\mu}(0) = 0$ ) the oscillations
are defined by the following expression:
$$
\nu_{e}(t) \cong \frac{1}{2} [ \exp(-iE'_{1}t) +
\exp(-iE'_{2}t)] \hspace{0.1cm}\nu_{e}(0)\\
$$
$$
\nu_{\mu}(t) \cong
\frac{1}{2} [ \exp(-iE'_{1}t) - \exp(-iE'_{2}t)] \hspace{0.1cm} \nu_{e}(0),
\eqno(6)
$$
or
$$
P(\nu_{e} \to \nu_{e}, t) = 1 - \sin^{2}(2\theta'(r)) \sin^{2}[\frac{\pi r}
{L'_{0}(r)}] ,\qquad r =ct ,
\eqno(7)
$$
(where $E'_{1}, E'_{2}$ are energies of $\nu_{1}, \nu_{2}$ neutrinos),
there neutrino resonance mixings take place.
\par
The equation (1) was obtained at supposition that the
neutrino behavior in the "matter" is analogous to a photon behavior
in matter with refraction coefficient-$n$.
\par
The photon velocity $c'$ in matter  with
refraction index $n$ is
$$
c' = \frac{c}{n} \eqno(8)
$$
and depends on characteristics of matter.
\par
The laws of conservation of the energy and the
momentum of the photon in such matter have the following form:
$$
E_{0} = E + E_{matt} = \frac{E_{0}}{n} + \frac{E_{0}(n - 1)}{n}
$$
$$
p_{0} = p + p_{matt} = \frac{p_{0}}{n} + \frac{p_{0}(n - 1)}{n}
\eqno(9)
$$
$$ E_{0} = \hbar w_{0}, w = \frac{w_{0}}{n}, E = pc' ,
$$
where $E_{0}, p_{0}$ are primary energy and momentum of the photon, $E, p,$
$E_{matt}, p_{matt}$ are, respectively, energy
and momentum of the photon in matter and energy and momentum of
the matter
polarization of the passing photon (matter response).
\par
If we suppose that the neutrinos in "matter" behave in analogy with the
photon in matter and the neutrino refraction indices are defined by the
expression
$$
n_{i} = 1 + \frac{2 \pi N}{p^{2}} f_{i}(0) ,
\eqno(10)
$$
(where $i$ is type of neutrinos
$(e, \mu, \tau)$, $N$ is density of "matter", $f_{i}(0)$ are a real part of
the forward scattering amplitude), then the velocity of neutrinos in "matter"
is determined by $n_{i}$.
\par
The electron neutrino ($\nu_{e}$)  in
"matter" interacts via $W^{\pm}, Z^{0}$ bosons and $\nu_{\mu}, \nu_{\tau}$
interact only via $Z^{0}$ boson. These differences in interactions lead to
the following differences on the refraction coefficients of $\nu_{e}$ and
$\nu_{\mu}, \nu_{\tau}$
$$
\Delta n = \frac{2 \pi N}{p^{2}} \Delta f(0) ,
\eqno(11)
$$
$$
\Delta f(0) = - \sqrt{2} \frac{G_F}{2 \pi} ,
$$
where $G_F$ is Fermi constant.
\par
Therefore the velocities of $\nu_{e}$ and $\nu_{\mu}, \nu_{\tau}$
in "matter" are different. And at the suitable density of "matter"
this difference can lead to a resonance enhancement of neutrino
oscillations in "matter" [4, 1].
\par
Since this considered hypothetical weak interaction is left-right
symmetric one, it is clear that we can use some analogy with
the Electrodynamics. Then
elastic and inelastic interactions can take place in matter.
Here we are
interested only in elastic interactions, namely potential interactions
of the charged particles in matter. These interactions lead
to polarization of the matter, in the result of it a definite part
of energy and momentum of the
massive charged particles go for  polarization of matter. The laws of
conservation of the energy and the momentum of the charged particle in
matter have the following form:
$$
E_{0} = E + E_{matt}
$$
$$
p_{0} = p + p_{matt},
\eqno(12)
$$
where $E_{0}, p_{0}$ are primary energy and momentum of the charged particle;
$E, p, E_{matt}, p_{matt}$ are, respectively, energy and momentum of
the charged particle in  matter and energy and momentum of
the matter polarization.
\par
It is clear that the matter polarization moves with the velocity
which is equal to the velocity of the charged particle in matter,
$v$, if this velocity $v$ is less than the velocity of light in
matter $c'$. If $v$ is equal or larger than
 $c'$, the energy and the momentum of polarization will go for the Cherenkov
radiation [5], i.e., the energy and the momentum losses will take place.
\par
It is interesting to know distributions of the energy and momentum
between the
charged particle and the matter polarization or which part of the
energy goes for mass alteration of the charged massive particle.
To solve this problem, it is necessary to
do a detailed computing of this interaction (connection between $m_{0},
m$ can be obtained using equations (12)).
Since it is out of our interest, we do not do this
computing. Our interest is connected with the problem
of resonance neutrino oscillations
in the hypothetical left-right symmetrical interaction which is used in
Wolfenstein's equation.
\par
So, from Wolfenstein's equation (1) with the hypothetical
left-right symmetric weak interaction we obtain the resonance neutrino
oscillation enhancement in "matter".
\par
One needs to notice, from the
analogy of the electrodynamics, that the considered process is an elastic
one, therefore and if even the neutrino resonance enhancement arises inside
the "matter" (the Sun), when these neutrinos go out from "matter" (the Sun)
into
vacuum, this enhancement disappears without leaving a trace and the
oscillations transit to vacuum neutrino oscillations (in vacuum the masses,
energies and momenta of the neutrinos are restored). The same result is
obtained from Eq. (6) since in vacuum  $ \frac{L'_{0}}{{L^{0}}} \to 0 $ and
$\sin^{2}(2\theta') \to  \sin^{2}(2\theta)$.
Hence, we come to conclusion that  behavior of neutrinos in the "matter"
is  like behavior of photons or charged particles (effective masses and
momenta are changed) in matter but not the behavior of vector polarization
of photons (i.e., rotation of photon vector polarization) in
matter as it is supposed in [4].
\par
It is very interesting to notice that
in the considered case (the "matter" with the hypothetical left-right
symmetrical weak interaction) if
$$
n_{i} -1 > 0 ,
\eqno(13)
$$
then, in analogy with the electrodynamics, when $v_{i} > \frac{c}{n_{i}} $,
the Cherenkov radiation will take place.

\par
Let us come to consideration of neutrinos passing
through real matter (i.e., in the case when only right components
of neutrinos participate in the weak interaction).

\section{Real Neutrinos in Matter or Impossibility to Realize the
Mechanism of Resonance Enhancement of Neutrino Oscillations in Matter
in the Framework of the Standard Weak Interaction}

\par
There is a question: can  real neutrinos in matter behave as
the photon or massive charged particle in matter? Then the
resonance enhancement of neutrino oscillations in matter will take place.
\par
Let us pass to a more detailed discussion of the problem of resonance
enhancement of
neutrino oscillations in matter through the standard weak interaction.
\par
As  is known (see Appendix and also [1, 3]), this interaction
cannot generate the masses.
Therefore, when the massive neutrino is passing through mater, its
mass does not change.
\par
The laws of conservation of the neutrino energy
and the momentum have the following form (we do not take into account
inelastic processes):
$$
a)\hspace{0.2cm} E_{0} = E +  W ,
$$
$$
b)\hspace{0.2cm} p_{0} = p + W \beta ,
\eqno(14)
$$
where $E_{0}, p_{0}, E, p$ are, respectively, energy and momentum neutrino in
vacuum and in matter, $W$ is elastic energy of neutrino interactions
in matter,
$\beta = \frac{v}{c}$.
\par
It is obvious that the response of matter moves with the velocity
$c (\beta = 1)$ since the weak interaction cannot generate a mass
(the mass of system does not change).
If we put expression b) into
expression a), we obtain the following expression:
$$
\sqrt{p_{0}^2 + m_{0}^2} = \sqrt{(p_{0} - W \beta)^{2} +
m_{0}^{2}} +  W ,
\eqno(15)
$$
which is solved only if:
$$
m_{0} = 0,
$$
or
$$
W = 0.
\eqno(16)
$$
The requirement of
conservations of energy and momentum for the weak interacting particle in
matter leads to a conclusion that or
$$
m_{0} = 0\qquad then \qquad W \neq 0,
$$
or if
$$
m_{0} \neq 0 \qquad then \qquad W = 0 ,
$$
it means that
the energy of matter polarization by neutrinos (or the energy of the
matter response $W$) must be zero $W = 0$, i.e., only
elastic and inelastic interactions of the neutrinos in matter take place
and there is no permanent response of matter (in contrast to
the Electrodynamics).
\par
As soon as the neutrinos are massive particles in Wolfenstein's equation,
$W $ must be equal to zero, and, this is why there
are no any changes of neutrino oscillations in matter.
\par
In the
reverse cases (when $m_{0i} = 0$) the $W$ can differ from zero, but in this
case, as it is well known, the vacuum oscillation of
neutrinos cannot take place.
\par
So, we come to a conclusion:
no resonance enhancement of neutrino oscillations in matter arises
through the standard weak interaction
\par
It is interesting to remark that, when the neutrinos are
passing through matter, there is no permanent polarization ($n_{i} = 1$)
of the matter, that is why the Cherenkov radiation cannot arise there.

\section{Conclusion}

In this paper, we studied a photon, a massive charged particle, and a
massive neutrino passing through matter.
The hypothetical left-right symmetric weak interaction, which is used
in Wolfenstein's
equations can generate the resonance enhancement of
neutrino oscillations in matter, which disappears when
neutrinos go out into vacuum from matter (the Sun).
\par
It was shown that
since standard weak interactions cannot generate masses, the laws of
conservation of the energy and the momentum of neutrino in matter will be
fulfilled only if the energy $W$ of polarization of matter by
the neutrino or the
corresponding term in Wolfenstein's equation, is zero.
This result implies that
neutrinos cannot generate permanent polarization of matter,  this
leads to the
conclusion:  resonance enhancement of neutrino oscillations in matter does
not exist.
\par
It was also shown that in standard weak interactions the
Cherenkov radiation cannot exist either.
\par
The nonresonance mechanism of the neutrino
oscillations enhancement in matter through the quasi-elastic weak
interactions was suggested in [6].
\par
In conclusion we would like to
stress that in the experimental data from [7] there is no visible change in
the spectrum of the $B^{8}$ Sun neutrinos. The measured spectrum of
neutrinos lies lower than the computed spectrum of the $B^{8}$
neutrinos [8]. In the case of realization of the resonance
enhancement  mechanism this spectrum must be distorted.
\par
The author expresses deep gratitude to Prof. A.A. Tyapkin
for the discussion of this work.

\section{Appendix}
We will operate in the framework of the quantum theory.
\par
Let $\hat M$ be a mass operator and $\Psi$- a wave function or state
function of a fermion. Then mass of the fermion is:
$$
M = <\Psi \mid \hat M \mid \Psi> .
\eqno(A.1)
$$
\par
If $\Psi$ is separated on the left and right states
$\Psi = \Psi_L + \Psi_R(\bar \Psi = \bar \Psi_L + \bar \Psi_R)$
then one can rewrite (A.1) in the following form:
$$
M = <\Psi_L \mid \hat M \mid \Psi_R> + <\Psi_R \mid \hat M \mid \Psi_L> .
\eqno(A.2)
$$
\par
Since the $\Psi_R (\bar \Psi_R)$ do not take part in weak interactions
of $W$ bosons, i.e. $\Psi_R = \bar \Psi_R \equiv 0$, then one obtains:
$$
M_w = <\Psi_L \mid \hat M \mid 0> + <0 \mid \hat M \mid \Psi_L> \equiv
0.
\eqno(A.3)
$$
\par
Let $\hat M^2$ is the squared mass operator .
Then the squared fermion mass is
$$
M^2 = <\Psi_L \mid \hat M^2 \mid
\Psi_R> + <\Psi_R \mid \hat M^2 \mid \Psi_L> .
\eqno(A.4)
$$
\par
Since the $\Psi_R (\bar \Psi_R)$ do not take part in weak
interactions of $W$ bosons, i.e. $\Psi_R = \bar \Psi_R \equiv 0$, then
one obtains:
$$
M^2_w \equiv 0.
\eqno(A.5)
$$
\par
REFERENCES
\par
\noindent
1. Kh.M. Beshtoev, JINR Communication E2-97-227, Dubna, 1997.
\par
See also the following works: Kh. M. Beshtoev, JINR Communication
\par
E2-91-183, Dubna, 1991;
\par
Proc. of 3rd Intern. Sym. on Weak and Electr. Int. in Nucl.,
\par
Dubna, 1992.
\par
\noindent
2. L. Wolfenstein, Phys.Rev. {\bf D 17}, 2369, (1978);
\par
Phys. Rev. {\bf D 20}, 2637, (1979).
\par
\noindent
3. Kh.M. Beshtoev, JINR Communication E2-93-167, Dubna, 1993;
\par
JINR Communication P2-93-44, Dubna, 1993;
\par
Fiz. Elem. Chastits At. Yadra, {\bf 27}, 53, (1996).
\par
\noindent
4. S.P. Mikheyev, A.Yu. Smirnov, Yad. Fiz. {\bf 42}, 1441, (1985);
\par
Sov. Phys. JETP {\bf 91}, 7, (1986);
\par
S.P. Mikheyev, A. Yu. Smirnov, Nuovo Cimento {\bf C 9}, 17, (1986);
\par
J. Boucher et al., Z. Phys. {\bf C 32}, 499, (1986).
\par
\noindent
5. I.M. Frank and I.E. Tamm, Dokl. Akad. Nauk USSR, {\bf 14}, 107,
\par
(1937).
\par
Kh.M. Beshtoev, JINR Communication E2-97-31, Dubna, 1997.
\par
\noindent
6. Kh.M. Beshtoev, Hadronic Journal, {\bf 18}, 165, (1995).
\par
\noindent
7. K.S. Harita, et al., Phys. Rev. Lett. {\bf65}, 1297, (1991);
\par
Phys. Rev., {\bf D 44}, 2341, (1991);
\par
Y. Totsuka, Rep. Prog. Physics 377, (1992).
\par
Y. Suzuki, Proceed. of the Intern. Conf. Neutrino-98, Japan, 1998.
\par
\noindent
8. J.N. Bahcall, Neutrino Astrophysics, Cambridge U.P.
\par
Cambridge, 1989.
\end{document}